\documentclass[11pt,twoside]{article}
\usepackage{./asp2014}

\aspSuppressVolSlug
\resetcounters

\begin{document}

\title{Solar ALMA: Observation-Based Simulations of the mm and sub-mm Emissions from Active Regions }
\author{Gregory Fleishman$^{1,2}$, Maria Loukitcheva$^{3,4}$, and Gelu Nita$^1$
\affil{$^1$Center For Solar-Terrestrial Research, New Jersey Institute of Technology, Newark, NJ 07102, USA; \email{gfleishm@njit.edu}; \email{gnita@njit.edu}}
\affil{$^2$Ioffe Institute, St. Petersburg 194021, Russia}
\affil{$^3$Astronomical Institute, St.Petersburg University, St.Petersburg, Russia; \email{marija@peterlink.ru}}
\affil{$^4$Max-Planck-Institut for Sonnensystemforschung, G\"ottingen, Germany; \email{lukicheva@mps.mpg.de}}}

\paperauthor{Sample~Author1}{Author1Email@email.edu}{ORCID_Or_Blank}{Author1 Institution}{Author1 Department}{City}{State/Province}{Postal Code}{Country}
\paperauthor{Sample~Author2}{Author2Email@email.edu}{ORCID_Or_Blank}{Author2 Institution}{Author2 Department}{City}{State/Province}{Postal Code}{Country}
\paperauthor{Sample~Author3}{Author3Email@email.edu}{ORCID_Or_Blank}{Author3 Institution}{Author3 Department}{City}{State/Province}{Postal Code}{Country}

\begin{abstract}
We developed an efficient algorithm integrated in our 3D modeling tool, GX Simulator \citep{Nita_etal_2015},
allowing quick computation of the synthetic {intensity and polarization} maps of solar active regions (AR) in the ALMA spectral range. 
\end{abstract}

To get an idea of how a given AR looks at the ALMA wavelengths we explore the recently updated set of 1D models (1D distributions of the electron temperature and density along with non-LTE ionized and neutral hydrogen densities with height) of the solar atmosphere \citep{2009ApJ...707..482F}, distinguishing between umbra, penumbra, network, internetwork, enhanced network, facula, or plage. 
Thus, to apply a given static atmospheric model to a given line of sight, the algorithm analyzes the photospheric input (white light and magnetogram) to classify the pixel as belonging to one of the itemized above photospheric features and creates a corresponding model mask (Fig.~\ref{GF_fig1}). Then, a 1D chromospheric model is added on top of each pixel, which forms a simplified 3D chromospheric model of the AR embedded in an extrapolated 3D magnetic data cube.
A huge advantage of this approach is that emission from any given AR can be synthesized very quickly, on the order of a few minutes after the AR selection.

Although simplified, these models are comprehensive enough to perform many tests; in particular, to compute anticipated averaged mm and sub-mm emission maps from a given AR as would be observed by ALMA. For the AR studies we specifically developed a computation unit taking accurately into account the gyroresonance (GR) and free-free processes with the full account of the magnetized plasma dispersion as well as the frequency-dependent mode coupling \citep{Fl_Kuzn_2014}.
Some results of this modeling for AR~12158 (observed 10 Sep 2014) are shown in Fig.~\ref{GF_fig2} for ALMA bands~3 and 6.
This modeling suggests that photospheric features will be distinguishable at the ALMA frequencies with the brightness temperature $T_{b}$ contrast up to a few hundred K. In line with the results of \citet{2014A&A...561A.133L}, the umbra, which is dark (cool) at the high frequencies, appears bright (hot) at the lower frequencies. In addition, $T_{b}$ noticeably depends on the magnetic field via its influence on the free-free opacity. Resulting degree of circular polarization reaches 3.6\% for AR~12158 (right panels in Fig.~\ref{GF_fig2}).
\vspace{-1in}
\articlefigure[width=.6\textwidth,bb=114 275 548 442,clip]{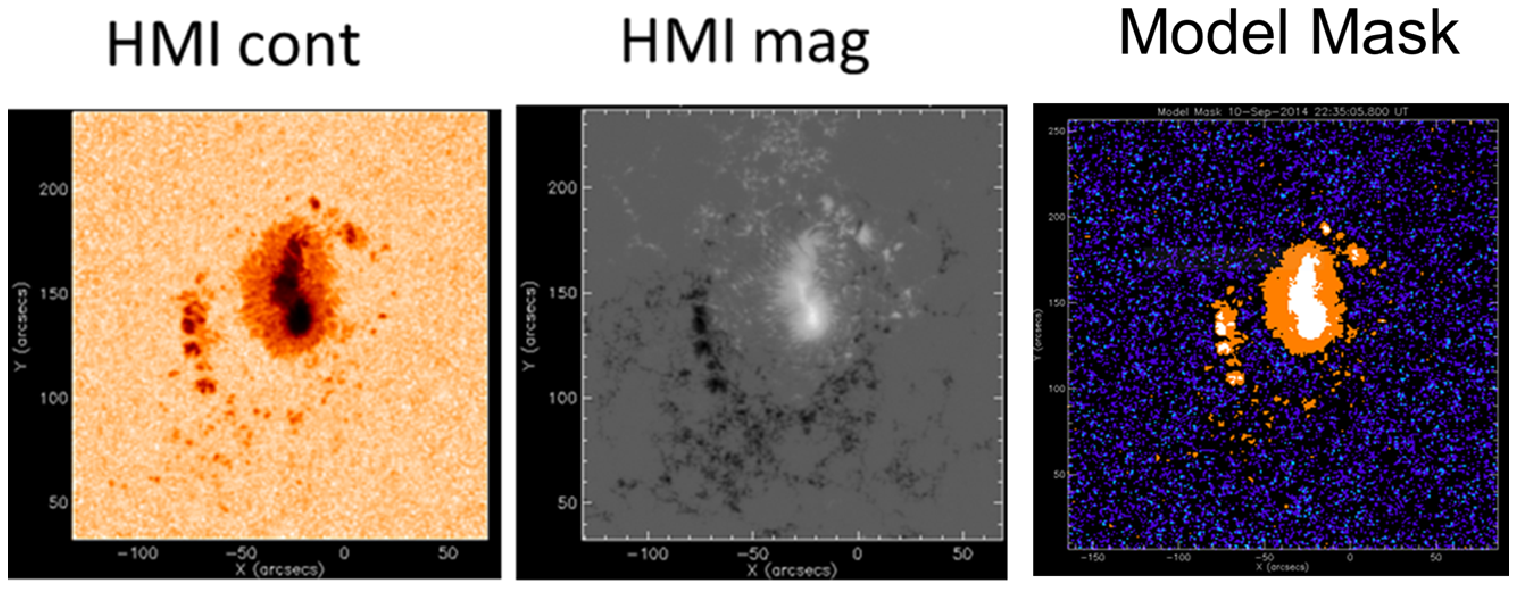}{GF_fig1}{{Left} and middle: Photospheric input to the chromospheric modeling of AR~12158: the white light image and magnetogram taken by SDO on Sep 10, 2014. {Right:} Model mask made based on the photospheric input.
}
\vspace{1in}
The degree of polarization along with the spatially resolved $T_{b}$ spectra are the keys in obtaining the magnetic field map at a given frequency, which translates to a certain chromospheric height. Multi-frequency data offer an elegant chromospheric tomography method (e.g., Loukitcheva et al., this volume). 
\articlefigure[width=.65\textwidth,clip]{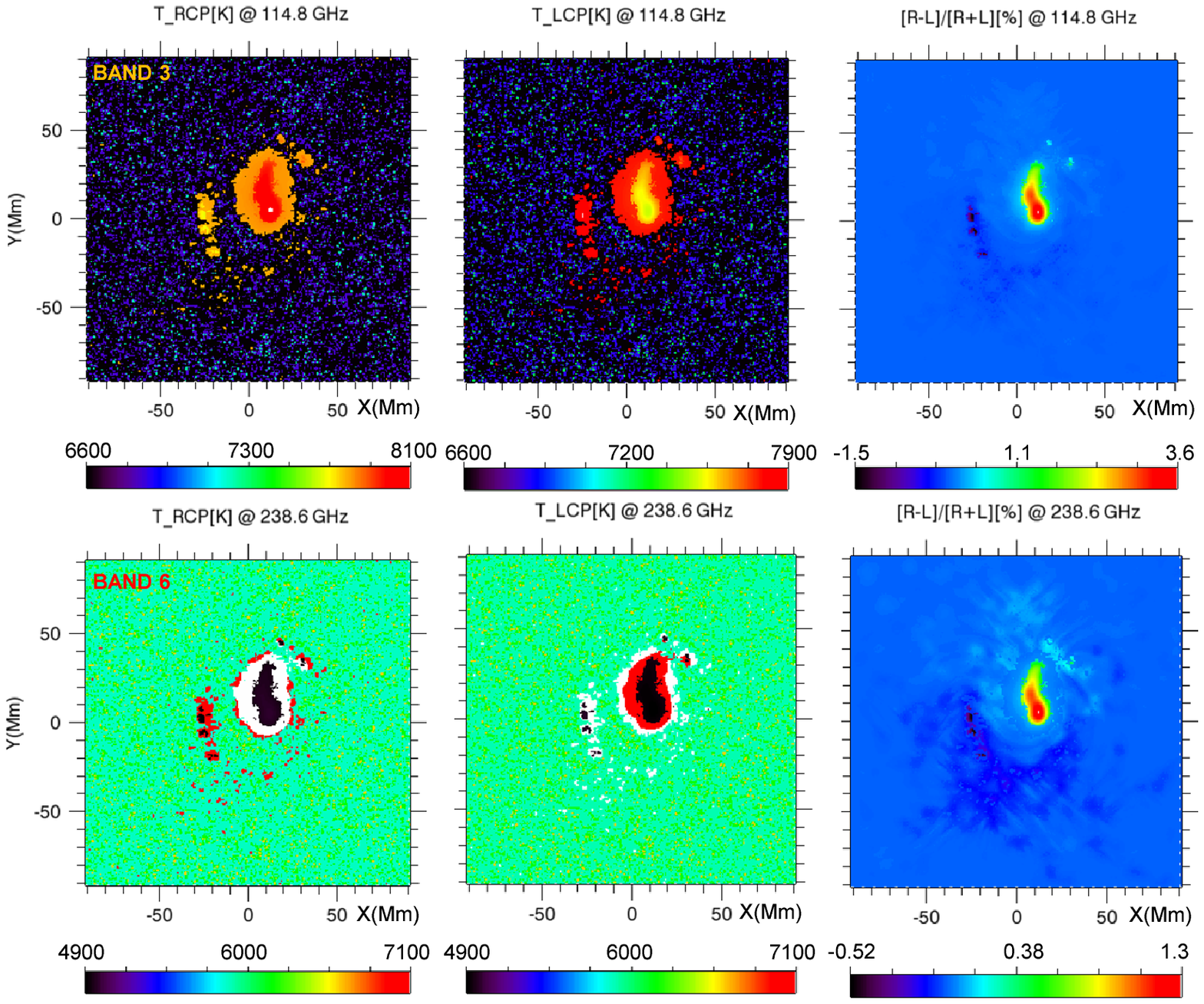}{GF_fig2}{Examples of simulated {LCP, RCP, and polarization} maps {in ALMA band~3 and band~6} for AR~12158. }

\acknowledgements This work was supported in part by NSF grants  AGS-1250374 and AGS-1262772, NASA grant NNX14AC87G, RFBR grants 15-02-01089, 15-02-03717, 15-02-03835, and 15-02-08028, and Saint-Petersburg State University research grants 6.0.26.2010 and 6.37.343.2015.

\bibliography{fleishman,chromo,loukitcheva}  





\end{document}